\documentclass[aps,prl,twocolumn,preprintnumbers,
             showpacs,nofootinbib]{revtex4}
\newcommand{\PRE}[1]{}       

\usepackage{bm} 
\usepackage{epsfig}

\newcommand{\kev}{\text{keV}}
\newcommand{\mev}{\text{MeV}}
\newcommand{\gev}{\text{GeV}}
\newcommand{\tev}{\text{TeV}}

\newcommand{\cm}{\text{cm}}

\newcommand{\s}{\text{s}}

\newcommand{\sr}{\text{sr}}

\newcommand{\eqref}[1]{Eq.~(\ref{#1})}

\newcommand{\figref}[1]{Fig.~\ref{fig:#1}}

\begin{document}

\preprint{UCI-TR-2007-17}

\title{
\PRE{\vspace*{1.5in}}
Resolving Cosmic Gamma Ray Anomalies with\PRE{\\}
Dark Matter Decaying Now\\
\PRE{\vspace*{0.3in}} }

\author{Jose~A.~R.~Cembranos}
\affiliation{Department of Physics and Astronomy, University of
California, Irvine, CA 92697, USA \PRE{\vspace*{.5in}} }
\author{Jonathan L.~Feng}
\affiliation{Department of Physics and Astronomy, University of
California, Irvine, CA 92697, USA \PRE{\vspace*{.5in}} }
\author{Louis E.~Strigari%
\PRE{\vspace*{.2in}} } \affiliation{Department of Physics and
Astronomy, University of California, Irvine, CA 92697, USA
\PRE{\vspace*{.5in}} }

\begin{abstract}
\PRE{\vspace*{0.3in}} Dark matter particles need not be completely
stable, and in fact they may be decaying now.  We consider this
possibility in the frameworks of universal extra dimensions and
supersymmetry with very late decays of WIMPs to Kaluza-Klein gravitons
and gravitinos.  The diffuse photon background is a sensitive probe,
even for lifetimes far greater than the age of the Universe.
Remarkably, both the energy spectrum and flux of the observed MeV
$\gamma$-ray excess may be simultaneously explained by decaying dark matter
with MeV mass splittings.  Future observations of continuum and line
photon fluxes will test this explanation and may provide novel
constraints on cosmological parameters.
\end{abstract}


\maketitle

The abundance of dark matter is now well known from observations of
supernovae, galaxies and galactic clusters, and the cosmic mocriwave
background (CMB)~\cite{Spergel:2006hy}, but its identity remains
elusive. Weakly-interacting massive particles (WIMPs) with weak-scale
masses $\sim 0.1 - 1~\tev$ are attractive dark matter candidates.  The
number of WIMPs in the Universe is fixed at freeze-out when they
decouple from the known particles about 1 ns after the Big Bang.
Assuming they are absolutely stable, these WIMPs survive to the
present day, and their number density is naturally in the right range
to be dark matter.  The standard signatures of WIMPs include, for
example, elastic scattering off nucleons in underground laboratories,
products from WIMP annihilation in the galaxy, and missing energy
signals at colliders~\cite{Bertone:2004pz}.

The stability of WIMPs is, however, not required to preserve the key
virtues of the WIMP scenario.  In fact, in supersymmetry (SUSY) and
other widely-studied scenarios, it is just as natural for WIMPs to
decay after freeze-out to other stable particles with similar masses,
which automatically inherit the right relic density to be dark
matter~\cite{Feng:2003xh}.  If the resulting dark matter interacts
only gravitationally, the WIMP decay is very late, in some cases leading to
interesting effects in structure formation~\cite{structureformation}
and other cosmological observables. Of course, the WIMP lifetime
depends on $\Delta m$, the mass splitting between the WIMP and its
decay product.  For high degeneracies, the WIMP lifetime may be of the
order of or greater than the age of the Universe $t_0 \simeq 4.3
\times 10^{17}~\s$, leading to the tantalizing possibility that dark
matter is decaying now.

\begin{figure}
\resizebox{2.61in}{!}{
\includegraphics{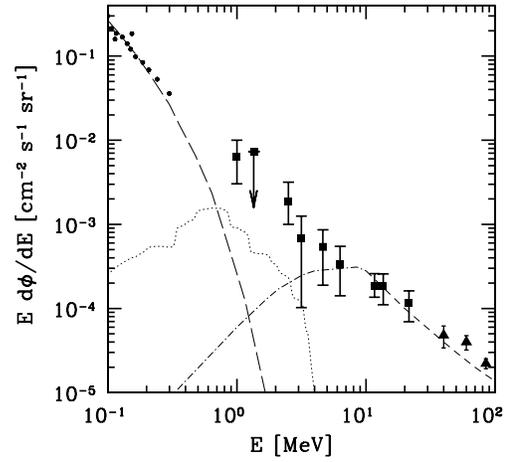}
} 
\caption{The CGB measured by HEAO-1~\cite{Gruber:1999yr} (circles),
COMPTEL~\cite{Weidenspointner} (squares), and
EGRET~\cite{Sreekumar:1997un} (triangles), along with the known
astrophysical sources: AGN (long-dash), SNIa (dotted), and blazars
(short-dash, and dot-dashed extrapolation).
\vspace*{-.2in}
\label{fig:cgbastrophysical1} }
\end{figure}

For very long WIMP lifetimes, the diffuse photon background is a
promising probe~\cite{Feng:2003xh,Ahn:2005ck}. Particularly
interesting is the (extragalactic) cosmic gamma ray background (CGB)
shown in \figref{cgbastrophysical1}.  Although smooth, the CGB
must be explained by multiple sources.  For $E_{\gamma} \alt 1~\mev$
and $E_{\gamma} \agt 10~\mev$, the CGB is reasonably well-modeled by
thermal emission from obscured active galactic nuclei (AGN)
\cite{Ueda:2003yx} and beamed AGN, or blazars~\cite{Pavlidou:2002va},
respectively. However, in the range $1~\mev \alt E_{\gamma} \alt
5~\mev$, no astrophysical source can account for the observed CGB.
Blazars are observed to have a spectral cut-off $\sim 10~\mev$, and
also only a few objects have been detected below this
energy~\cite{McNaron-Brown,Stecker:1999hv}; a maximal upper
limit~\cite{Comastri} on the blazar contribution for $E_{\gamma} \alt
10~\mev$ is shown in \figref{cgbastrophysical1}.  Diffuse $\gamma$-rays from 
Type Ia supernovae (SNIa) contribute below $\sim 5~\mev$, but the most 
recent astronomical data show that they also cannot account for the entire 
spectrum~\cite{Strigari:2005hu,Ahn:2005ws}; previous calculations suggested 
that SNIa are the dominant source of $\gamma$-rays at MeV energies
\cite{SNIaprevious}.

In this paper, we study the contribution to the CGB from dark matter
decaying now.  We consider simple models with extra dimensions or SUSY
in which WIMP decays are highly suppressed by both the weakness of
gravity and small mass splittings and are dependent on a single
parameter, $\Delta m$.  We find that the CGB is an extremely sensitive
probe, even for lifetimes $\tau \gg t_0$.  Intriguingly, we also find
that both the energy spectrum and the flux of the gamma ray excess
described above are naturally explained in these scenarios with
$\Delta m \sim \mev$.

As our primary example we consider minimal universal extra dimensions
(mUED)~\cite{Appelquist:2000nn}, one of the simplest imaginable models
with extra dimensions.  In mUED all particles propagate in one extra
dimension compactified on a circle, and the theory is completely
specified by $m_h$, the Higgs boson mass, and $R$, the
compactification radius. (In detail, there is also a weak, logarithmic
dependence on the cutoff scale $\Lambda$~\protect\cite{Cheng:2002iz}.
We present results for $\Lambda R=20$.)  Every particle has a
Kaluza-Klein (KK) partner at every mass level $\sim m/R$, $m=1, 2,
\ldots$, and the lightest KK particle (LKP) is a dark matter
candidate, with its stability guaranteed by a discrete parity.

Astrophysical and particle physics constraints limit mUED parameters
to regions of $(R^{-1}, m_h)$ parameter space where the two lightest
KK particles are the KK hypercharge gauge boson $B^1$, and the KK
graviton $G^1$, with mass splitting $\Delta m \alt {\cal
O}(\gev)$~\cite{Cembranos:2006gt}. This extreme degeneracy, along with
the fact that KK gravitons interact only gravitationally, leads to
long NLKP lifetimes
\begin{eqnarray}
\tau& \simeq &
\frac{3\pi}{b \cos^2 \theta_W} 
\frac{M_P^2}{(\Delta m)^3}\simeq
\frac{4.7 \times 10^{22}~\s}{b}
\left[\frac{\mev}{\Delta m} \right]^3 \, ,
\label{gravidecay}
\end{eqnarray}
where $M_P \simeq 2.4 \times 10^{18}~\gev$ is the reduced Planck
scale, $\theta_W$ is the weak mixing angle, $b= 10/3$ for $B^1 \to G^1
\gamma$, and $b= 2$ for $G^1 \to B^1 \gamma$~\cite{Feng:2003nr}.  Note
that $\tau$ depends only on the single parameter $\Delta m$.  For
$795~\gev \alt R^{-1} \alt 809~\gev$ and $180~\gev \alt m_h \alt
215~\gev$, the model is not only viable, but the $B^1$ thermal relic
abundance is consistent with that required for dark
matter~\cite{Kakizaki:2006dz} and $\Delta m \alt 30~\mev$, leading to 
lifetimes $\tau ( B^1 \to G^1 \gamma) \agt t_0$.

We will also consider supersymmetric models, where small mass
splittings are also possible, since the gravitino mass is a completely
free parameter.  If the two lightest supersymmetric particles are a
Bino-like neutralino $\tilde{B}$ and the gravitino $\tilde{G}$, the
heavier particle's decay width is again given by \eqref{gravidecay},
but with $b=2$ for $\tilde{B} \to \tilde G \gamma$, and $b=1$ for
$\tilde{G} \to \tilde{B} \gamma$.  As in mUED, $\tau$ depends only on
$\Delta m$, and $\Delta m \alt 30~\mev$ yields lifetimes greater than
$t_0$.

The present photon flux from two-body decays is
\begin{equation}
\frac{d\Phi}{ dE_{\gamma}}
=\frac{c}{4\pi} \int_0^{t_0} \frac{dt}{\tau}
\frac{N(t)}{V_0}
\delta \left( E_{\gamma} - a \varepsilon_\gamma \right) ,
\label{differentialflux}
\end{equation}
where $N(t) = N^{\text{in}} e^{-t/\tau}$ and $N^{\text{in}}$ is the
number of WIMPs at freeze-out, $V_0$ is the present volume of the
Universe, $a$ is the cosmological scale factor with $a(t_0)\equiv 1$,
and $\varepsilon_\gamma = \Delta m$ is the energy of the produced
photons.  
Photons from two-body decays are observable in the diffuse photon
background only if the decay takes place in the late Universe, when
matter or vacuum energy dominates.  In this case,
\eqref{differentialflux} may be written as
\begin{equation}
\frac{d\Phi}{ dE_{\gamma}} =
\frac{c}{4\pi} \frac{N^{\text{in}}\,
e^{-P(E_\gamma /\varepsilon_{\gamma})/\tau}}
{V_0 \tau E_{\gamma} H(E_\gamma /\varepsilon_{\gamma})} 
 \, \Theta(\varepsilon_\gamma - E_{\gamma}) \,, 
\label{extragalacticphi}
\end{equation}
where $P(a)= t$ is the solution to $(da/dt)/a = H(a)= H_0
\sqrt{\Omega_M a^{-3} + \Omega_{\text{DE}} \, a^{-3(1+w)}}$ with
$P(0)=0$, and $\Omega_M$ and $\Omega_{\text{DE}}$ are the matter and
dark energy densities. If dark energy is a cosmological constant
$\Lambda$ with $w=-1$,
\begin{equation}
P(a)\equiv \frac{2 \ln \left[ \left(\sqrt{\Omega_{\Lambda}a^3}
+\sqrt{\Omega_{M}+\Omega_{\Lambda}a^3}\right) / 
\sqrt{\Omega_{M}} \, \right]}
{3 H_0 \sqrt{\Omega_{\Lambda}}}\,. 
\label{P}
\end{equation}
The flux has a maximum at 
$E_{\gamma}= \varepsilon_{\gamma}
[\frac{\Omega_{M}}{2\Omega_{\Lambda}}
U(H_0^2\tau^2\Omega_{\Lambda})]^{\frac{1}{3}}$,
where $U(x)\equiv (x+1-\sqrt{3x+1})/(x-1)$.

The energy spectrum is easy to understand for very long and very short
decay times.  For $\tau \ll t_0$, $H_0^2\tau^2\Omega_{\text{DE}}\ll
1$, and the flux grows due to the decelerated expansion of the
Universe as $d\Phi/dE_{\gamma} \propto E^{1/2}$ until it reaches its
maximum at $E^{\text{max}}_{\gamma}\simeq \varepsilon_{\gamma}
(\Omega_{M}H_0^2\tau^2/4)^{1/3}$.  Above this energy, the flux is
suppressed exponentially by the decreasing number of decaying
particles~\cite{Feng:2003xh}.

On the other hand, if $\tau \gg t_0$, $H_0^2 \tau^2 \Omega_{\text{DE}}
\gg 1$, and the flux grows as $d\Phi/dE_{\gamma} \propto E^{1/2}$ only
for photons that originated in the matter-dominated epoch. For decays
in the vacuum-dominated Universe, the flux decreases asymptotically as
$d\Phi/dE_{\gamma} \propto E^{(1+3w)/2}$ due to the accelerated
expansion.  The flux reaches its maximal value at
$E^{\text{max}}_{\gamma}\simeq \varepsilon_{\gamma}
[-\Omega_{M}/((1+3w)\Omega_{\text{DE}})]^{-1/(3w)}$ where photons were
produced at matter-vacuum equality. Note that this value and the
spectrum shape depend on the properties of the dark energy.  Assuming
$\Omega_M = 0.25$, $\Omega_{\text{DE}} = 0.75$, $w=-1$, and $h = 0.7$,
and that these particles make up all of non-baryonic dark matter, so
that
\begin{equation}
\frac{N^{\text{in}}}{V_0} = 1.0 \times 10^{-9}~\cm^{-3}
\left[\frac{\tev}{m} \right]
\left[\frac{\Omega_{\text{NBDM}}}{0.2} \right] \, ,
\label{nbdm}
\end{equation}
we find that the maximal flux is
\begin{eqnarray}
&&\frac{d\Phi}{dE_{\gamma}} (E_\gamma^{\text{max}})
= 1.33\times 10^{-3}~\cm^{-2}~\s^{-1}~\sr^{-1}~\mev^{-1} \nonumber \\
&&\qquad \times \left[ \frac{\tev}{m} \right]
\left[ \frac{\mev} {\Delta m}\right]
\left[ \frac{10^{21}~\s} {\tau}\right]^{\frac{2}{3}}
\left[ \frac{\Omega_{\text{NBDM}}}{0.2} \right] \, .
\label{phimax}
\end{eqnarray}

\begin{figure}
\resizebox{2.61in}{!}{
\includegraphics{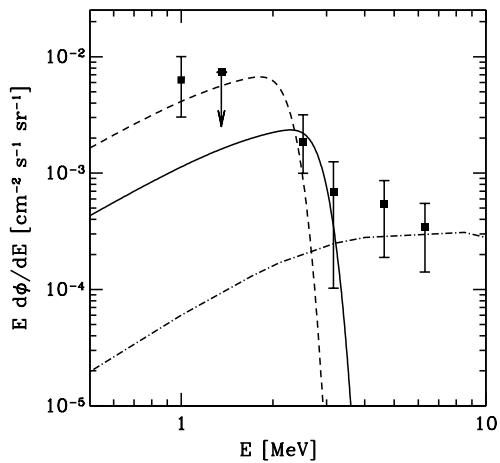}
} 
\caption{Data for the CGB in the range of the MeV excess, along with
predicted contributions from extragalactic dark matter decay.  The
curves are for $B^1 \to G^1 \gamma$ in mUED with lifetime $\tau =
10^3\, t_0$ and $m_{B^1}=800~\gev$ (solid) and $\tilde{B} \to \tilde G
\gamma$ in SUSY with lifetime $\tau = 5 \times 10^3\, t_0$ and
$m_{\tilde{B}}=80~\gev$ (dashed).  We have assumed
$\Omega_{\text{NBDM}} = 0.2$ and smeared all spectra with energy
resolution $\Delta E/E = 10\%$, characteristic of experiments such as
COMPTEL. The dot-dashed curve is the upper limit to the blazar spectrum, 
as in Fig.~(\ref{fig:cgbastrophysical1}). 
\vspace*{-.2in}
\label{fig:cgb} }
\end{figure}

\figref{cgb} shows example contributions to the CGB 
from decaying dark matter in mUED and SUSY.  The
mass splittings have been chosen to produce maximal fluxes at
$E_{\gamma} \sim \mev$. These frameworks are, however, highly
constrained: once $\Delta m$ is chosen, $\tau$ and the flux are
essentially fixed.  It is thus remarkable that the predicted flux
is in the observable, but not excluded, range and may explain the
current excess above known sources.

\begin{figure}
\resizebox{2.61in}{!}{
\includegraphics{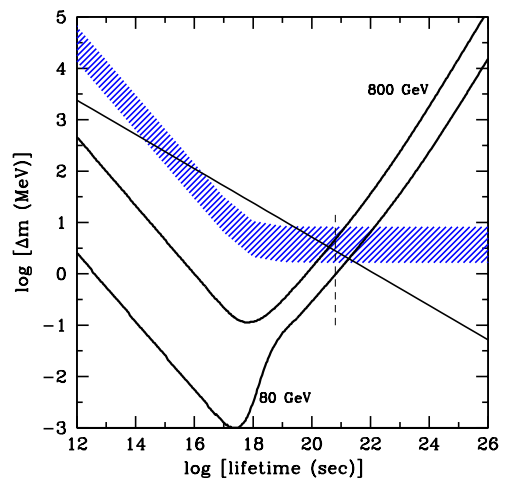}
} 
\caption{Model-independent analysis of decaying dark matter in the
$(\tau, \Delta m)$ plane.  In the shaded region, the resulting
extragalactic photon flux peaks in the MeV excess range $1~\mev \le
E^{\text{max}}_{\gamma} \le 5~\mev$.  On the contours labeled with
WIMP masses, the maximal extragalactic flux matches the extragalactic
flux observed by COMPTEL; points above these contours are excluded.
The diagonal line is the predicted relation between $\tau$ and $\Delta
m$ in mUED. On the dashed line, the predicted Galactic flux matches
INTEGRAL's sensitivity of $10^{-4}~\cm^{-2}~\s^{-1}$ for monoenergetic
photons with $E_{\gamma} \sim ~\mev$.
\vspace*{-.2in}
\label{fig:parameterspace} }
\end{figure}

To explore this intriguing fact further, we relax model-dependent
constraints and consider $\tau$ and $\Delta m$ to be independent
parameters in \figref{parameterspace}.  The labeled curves give the
points in $(\tau, \Delta m)$ parameter space where, for the WIMP
masses indicated and assuming \eqref{nbdm}, the maximal flux from
decaying dark matter matches the flux of the observed photon
background in the keV to 100 GeV range~\cite{Gruber:1999yr}.  For a
given WIMP mass, all points above the corresponding curve predict peak
fluxes above the observed diffuse photon background and so are
excluded.

The shaded band in \figref{parameterspace} is the region where the
maximal flux falls in the unaccounted for range of 1-5 MeV.  For
$\tau \agt t_0$, $E^{\text{max}}_{\gamma}\simeq 0.55\,\Delta
m$. However, for $\tau \alt t_0$, $E^{\text{max}}_{\gamma}$ does not
track $\Delta m$, as the peak energy is significantly redshifted. For
example, for a WIMP with mass 80 GeV, $\tau \sim 10^{12}~\s$ and
$\Delta m \sim \mev$, $E^{\text{max}}_{\gamma} \sim \kev$.  The
overlap of this band with the labeled contours is where the observed
excess may be explained through WIMP decays.  We see that it requires
$10^{20}~\s \alt \tau \alt 10^{22}~\s$ and $1~\mev \alt \Delta m \alt
10~\mev$.  These two properties may be simultaneously realized by
two-body gravitational decays: the diagonal line shows the relation
between $\tau$ and $\Delta m$ given in \eqref{gravidecay} for $B^1
\to G^1 \gamma$, and we see that this line passes through the overlap
region!  Similar conclusions apply for all other decay models
discussed above.

\begin{figure}
\resizebox{2.61in}{!}{
\includegraphics{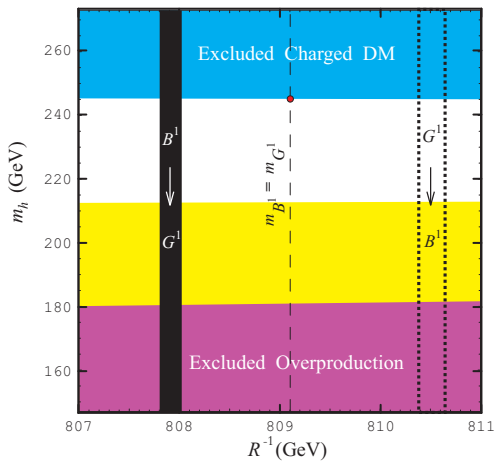}
} 
\caption{Phase diagram of mUED.  The top and bottom shaded regions are
excluded for the reasons indicated~\cite{Cembranos:2006gt}.  In the
yellow (light) shaded region, the $B^1$ thermal relic density is in
the $2\sigma$ preferred region for non-baryonic dark
matter~\cite{Kakizaki:2006dz}.  In the vertical band on the left
(right) the decay $B^1 \to G^1 \gamma$ ($G^1 \to B^1 \gamma$) can
explain the MeV diffuse photon excess.
\vspace*{-.2in}
\label{fig:muedspace} }
\end{figure}

These considerations of the diffuse photon background also have 
implications for the underlying models. For mUED, $\Delta m =
2.7-3.2~\mev$ and $\tau = 4-7 \times 10^{20}~\s$ can explain the MeV
excess in the CGB.  This preferred region is realized for the decay
$B^1 \to G^1 \gamma$ for $R^{-1} \approx 808~\gev$. (See
\figref{muedspace}.) Lower $R^{-1}$ predicts larger $\Delta m$ and
shorter lifetimes and is excluded. The MeV excess may also be realized
for $G^1 \to B^1 \gamma$ for $R^{-1}\approx 810.5~\gev$, though
in this case the $G^1$ must be produced
non-thermally to have the required dark matter
abundance~\cite{Feng:2003nr,Shah:2006gs}.

So far we have concentrated on the cosmic, or extragalactic, photon
flux, which is dependent only on cosmological parameters.  The
Galactic photon flux depends on halo parameters and so is less robust,
but it has the potential to be a striking signature, since these
photons are not redshifted and so will appear as lines with
$E_{\gamma} = \Delta m$.  
INTEGRAL has searched for photon lines  
within $13^{\circ}$ from the Galactic center~\cite{Teegarden:2006ni}. For lines with 
energy $E \sim \mev$ and width $\Delta E \sim 10~\kev$, INTEGRAL's energy resolution
at these energies, INTEGRAL's sensitivity is $\Phi \sim
10^{-4}~\cm^{-2}~\s^{-1}$.  The Galactic flux from decaying dark
matter saturates this limit along the vertical line in
\figref{parameterspace}, assuming $m_{\chi} = 800~\gev$.  This flux is
subject to halo uncertainties; we have assumed the halo density
profiles of Ref.~\cite{Klypin:2001xu}, which give a
conservative upper limit on the flux within the field of
view. Remarkably, however, we see that the vertical line also passes
through the overlap region discussed above.  If the MeV CGB anomaly is
explained by decaying dark matter, then, the Galactic flux is also
observable, and future searches for photon lines will stringently test
this scenario.

In conclusion, well-motivated frameworks support the possibility that
dark matter may be decaying now.  We have shown that the diffuse
photon spectrum is a sensitive probe of this possibility, even for
lifetimes $\tau \gg t_0$. This is the leading probe of these
scenarios. Current bounds from the CMB~\cite{Ichiki:2004vi} and
reionization~\cite{Chen:2003gz} do not exclude this scenario, but they
may also provide complementary probes in the future.  We have also
shown that dark matter with mass splittings $\Delta m \sim~\mev$ and
lifetimes $\tau \sim 10^3 - 10^4$ Gyr can explain the current excess
of observations above astrophysical sources at $E_{\gamma} \sim
\mev$. Such lifetimes are unusually long, but it is remarkable that
these lifetimes and mass splittings are simultaneously realized in
simple models with extra dimensional or supersymmetric WIMPs decaying
to KK gravitons and gravitinos.  Future experiments, such as
ACT~\cite{Boggs:2006mh}, with large apertures and expected energy
resolutions of $\Delta E/E = 1\%$, may exclude or confirm this
explanation of the MeV excess through both continuum and line signals.
Finally, we note that if dark matter is in fact decaying now, the
diffuse photon signal is also sensitive to the recent expansion
history of the Universe.  For example, as we have seen, the location
of the spectrum peak is a function of $\Omega_M/ \Omega_{\text{DE}}$
and $w$.  The CGB may therefore, in principle, provide novel
constraints on dark energy properties and other cosmological
parameters.

\begin{acknowledgments}
We thank John Beacom, Matt Kistler, and Hasan Yuksel for Galactic flux
insights.  The work of JARC and JLF is supported in part by NSF Grants
PHY--0239817 and PHY--0653656, NASA Grant No.~NNG05GG44G, and the
Alfred P.~Sloan Foundation. The work of JARC is also supported by the
FPA 2005-02327 project (DGICYT, Spain). LES and JARC are supported by
the McCue Postdoctoral Fund, UCI Center for Cosmology.
\end{acknowledgments}


\end{document}